\documentclass[12pt,preprint]{emulateapj}

\usepackage{graphicx}
\usepackage{amsmath}

\begin{document}

\title{The wave nature of continuous gravitational waves from microlensing}
\author{Kai Liao$^{1}$, Marek Biesiada$^{2,3}$, Xi-Long Fan$^{4,5*}$}
\affil{
$^1$ {School of Science, Wuhan University of Technology, Wuhan 430070, China.}\\
$^2$ {Department of Astronomy, Beijing Normal University, Beijing 100875, China.}\\
$^3$ {Department of Astrophysics and Cosmology, Institute of Physics, University of Silesia, 75 Pu{\l}ku Piechoty 1, 41-500 Chorz{\'o}w, Poland.}\\
$^4$ {School of Physics and Technology, Wuhan University, Wuhan 430072, China.}\\
$^5$ {Department of Physics and Mechanical and Electrical Engineering, Hubei University of Education, Wuhan 430205, China.}
}
\email{*fanxilong@outlook.com}

\begin{abstract}
Gravitational wave predicted by General Relativity is the transverse wave of spatial strain. Several gravitational
waveform signals from binary black holes and from a binary neutron star system accompanied by electromagnetic counterparts have been
recorded by advanced LIGO and advanced Virgo. In analogy to light, the spatial fringes of diffraction and interference should also exist
as the important features of gravitational waves.
We propose that observational detection of such fringes could be achieved through gravitational lensing of continuous gravitational waves. The lenses would play the role of the diffraction barriers. Considering peculiar motions of the observer, the lens and the source, the spatial amplitude variation of diffraction or interference fringes should be detectable as an amplitude modulation of monochromatic gravitational signal.
\end{abstract}
\keywords{gravitational lensing: micro - gravitational waves}

\section{Introduction}
Historically, the nature of light has been described as either wave or particles (corpuscules).
These different views have been fiercely debated and supported by famous experiments.
For example, Young's double-slit experiment, Kirchoff-Fresnel and Fraunhoffer's diffraction supported the wave nature of light, while
Einstein's photoelectric effect and Compton scattering suggested its corpuscular nature. Eventually, with the development of quantum mechanics
we accepted the wave-particle duality inherent to fundamental constituents of the world.
On the other hand, gravitational waves (GWs), predicted by the General Relativity (GR), are transverse waves of spatial strain, powered by varying in time quadruple moments of the source mass.
Reasoning by analogy, one often conjectures that like the photon,
the graviton should exist in the quantum theory of gravity, as a spin 2 massless particle traveling at the speed of light.
Recent successful detections of GW signals from astrophysical sources confirmed their classical nature.
However, in order to gain better empirical understanding of the physical nature of GW, in particular to better investigate its wave nature, it is essential to design appropriate further experiments or observation strategies. This is the main motivation behind this work.

GW signals from coalescing binary black holes~\citep{Abbott2016a,Abbott2016b,Abbott2017a} and a merger of binary neutron stars
~\citep{Abbott2017b} have been detected by LIGO and Virgo collaboration, giving birth to GW astronomy and rising multimessenger astronomy to the
new level. Theoretical chirp waveforms have been fitted to the observed signals leading to reliable estimates of physical characteristics of coalescing sources, like their chirp masses or luminosity distances. However, one should not forget that this happened only in one particular place (at Earth) and it would be
rewarding to register spatial fringes of the GW (as it was done for light) further confirming its wave nature. We propose an experimental setting to achieve this goal.

Gravitational lensing is another effect predicted by the GR~\citep[e.g. see a review and a history of  gravitational lensing  in][]{Treu2010,Sauer2007}, where the light traveling along null geodesics bends in the vicinity of massive bodies. Strong lensing by galaxies and
galaxy clusters, weak lensing by dark matter halos, microlensing by stars and millilensing by dark matter substructure
have been widely used in astrophysics~\citep{Zackrisson2010} and cosmology~\citep{Bonvin2017}. Similarly, GW also travels along the null geodesic and it should also display lensing effects.
If the wavelength of the GW is much shorter than the lens mass scale (like in the case of light), it behaves according to the geometric optics limit based
on Fermat's principle. In such case, gravitational lensing could affect the accuracy of physical inference made from registered signals~\citep{Oguri2018}.
The event rates of GW lensing by galaxies were studied by~\citep{Wang1996,Sereno2010,Biesiada2014,Ding2015,Li2018} leading to a robust predicition that third generation of GW interferometric detectors would yield $50 - 100$ lensed GW events per year.
Strongly lensed GW signals, observed together with their EM counterparts have been demonstrated to enhance our understanding regarding fundamental physics~\citep{Fan2017,Collett2017}, astrophysics~\citep{Liao2018} and cosmology~\citep{Sereno2011,Liao2017,Wei2017,Li2019}.
It is noteworthy that a very recent work showed the events GW170809 and GW170814 could come from the same source~\citep{Broadhurst2019}, but see \cite{Hannuksela2019}

On the contrary, if the wavelength is much longer than the lens mass scale (i.e. its Schwarzschild radius), then one should use the wave optics limit.
In the intermediate regime, wave superposition effects should be considered~\citep{Ohanian1974,Bliokh1975}. The lens acts like a diffraction barrier.
Wave effects in gravitational lensing of gravitational waves and their influence on the waveforms of chirp signals from inspiraling binaries were discussed in~\citep{Nakamura1998,Takahashi2003,Takahashi2017}.
Fringes due to GW lensing by compact dark matter~\citep{Jung2017}, stars~\citep{Christian2018} and intermediate-mass black hole~\citep{Lai2018}
were claimed to be detectable, while the waveforms of strongly lensed GWs were studied in~\citep{Dai2017}.
These works focused on the distortion of chirping waveforms as a function of  frequency. Therefore, they represent a waveform dependent approach to observe the wave effects of GWs and manifest themselves as systematic uncertainties in the parameter estimation~\citep{Cao2014}. Remarkably, \citet{Dai2018} proposed recently an agnostic detection method of diffraction effects on lensed chirp signals based on dynamic programming, which does not require a detailed model of the lensed waveforms. 

We propose a strategy to observe spatial diffraction or interference fringes directly, as the
spatial variations of the GW amplitude. We find that this could be achieved through monitoring the lensed source of continuous, periodic  GW for a time scale of
0.1-10 years due to relative peculiar motions of the strong lensing system components.
Throughout this paper we use natural units $c=G=1$ in all equations, SI units are recovered for the purpose of numerical estimates.

\section{Wave optics description}
\begin{figure*}
 \includegraphics[width=5cm,angle=0]{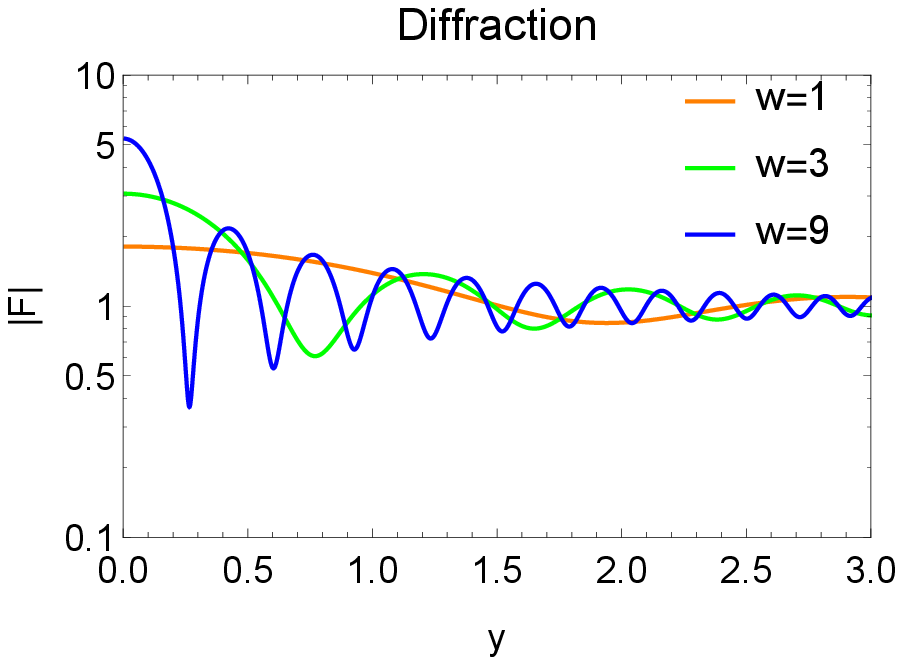}
 \includegraphics[width=5cm,angle=0]{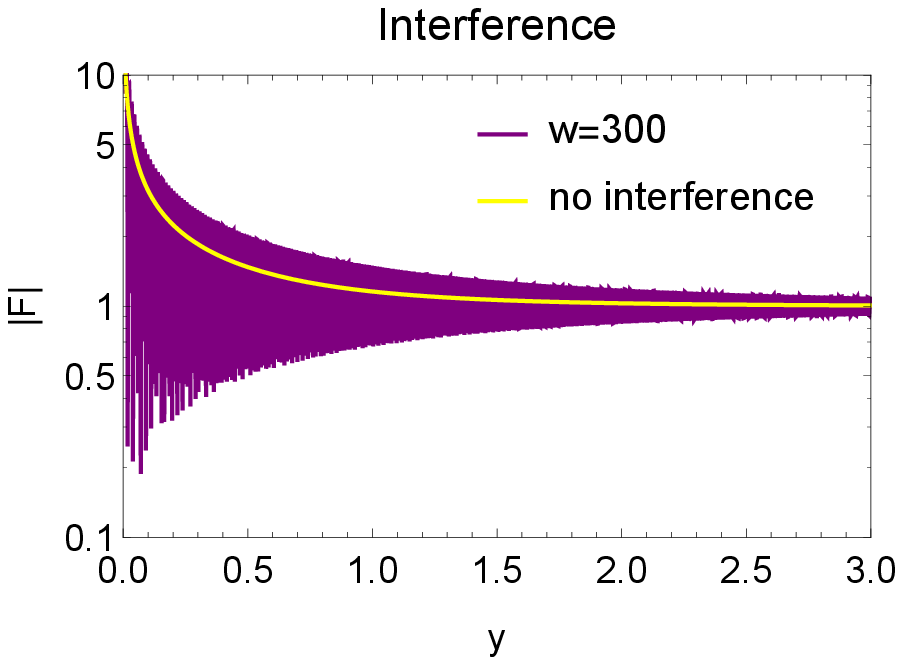}
 \includegraphics[width=5cm,angle=0]{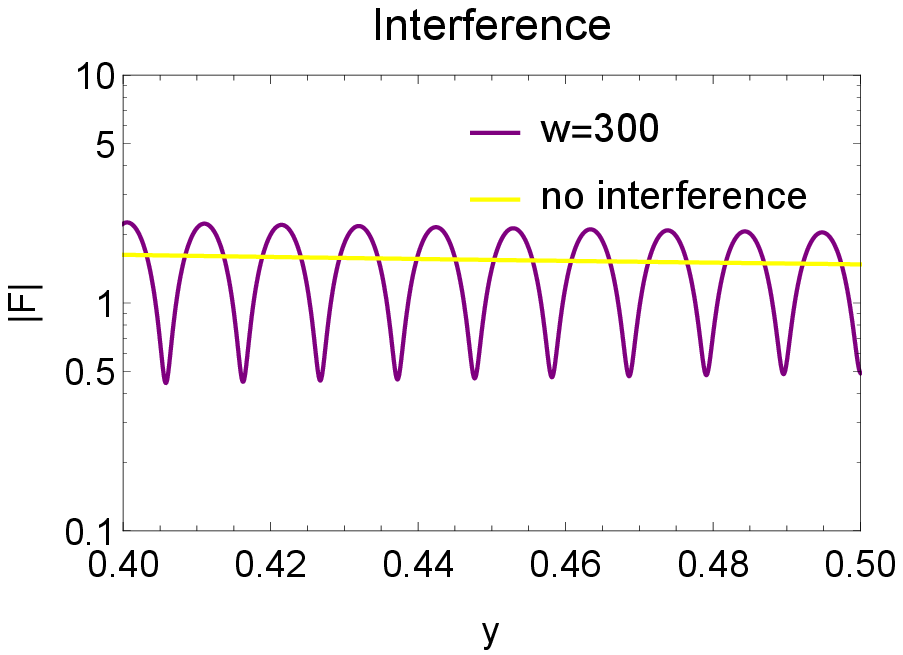}
   \caption{The relative amplitude of lensed continuous gravitational wave as a function of source position for point mass lens. The left: diffraction in wave optics description for different values of $w$ parameter. The middle: interference pattern of images in the geometric optics limit $w=300$. The right: zoom of the interval $y\in [0.4, 0.5]$. In the geometric optics limit, amplification of light from lensed images is shown in yellow for comparison.
  }\label{point}
\end{figure*}
In order to study the wave effects of GW due to lensing we consider GW propagating in the weak gravitational field of the lens (see more discussion on strong gravitational field in \cite{Zhang2018}). Thus, the metric can be written as:
\begin{equation}
g_{\mu \nu}=g^{(L)}_{\mu \nu}+h_{\mu \nu},
\end{equation}
where $g^{(L)}_{\mu \nu}$ is the metric determined by the Newtonian potential $U$ of the lens, $h_{\mu \nu}$ is the GW perturbation.
Since the polarization tensor $e_{\mu \nu}$ is parallel transported along null geodesic, we consider a
scalar wave:
\begin{equation}
h_{\mu \nu}=\phi e_{\mu \nu},
\end{equation}
whose propagation equation is determined by:
\begin{equation}
\partial_\mu(\sqrt{-g^{(L)}}g^{(L)\mu \nu}\partial_\nu\phi)=0.
\end{equation}
In the weak field limit and switching to the frequency ($f$) domain of this scalar wave $\tilde{\phi}(f,\mathbf{r})$ (  with $\dot{f} \ll 1$ ), one is able to rewrite the above equation in the form of Helmholtz equation:
\begin{equation}
(\Delta + 4\pi^2 f^2)\tilde{\phi}=16\pi^2f^2U\tilde{\phi}.
\end{equation}
Following~\cite{Takahashi2003}, we define the dimensionless amplification factor:
\begin{equation}
F(f)=\tilde{\phi}^L(f)/\tilde{\phi}(f),
\end{equation}
where $\tilde{\phi}^L$ and $\tilde{\phi}$ are the lensed and unlensed (U=0) amplitudes, respectively.

Diffraction results are as a superposition of all possible waves on the lens plane that have different time delays corresponding to different phases.
The amplification factor $F(f)$ is given by~\citep{Takahashi2003}:
\begin{equation}
F(f)=\frac{D_sR^2_E(1+z_l)}{D_lD_{ls}}\frac{f}{i}\int d^2\mathbf{x}\ exp\left[2\pi ift_d(\mathbf{x},\mathbf{y})\right]\label{amplitude},
\end{equation}
where $R_E$ is the Einstein radius of the lens, and by $\mathbf{x}=\mathbf{\xi}/R_E, \mathbf{y}=\mathbf{\eta} D_L/(R_ED_S)$, we denote dimensionless positions (normalized by $R_E$) with  $\mathbf{\xi}$ being the impact parameter in the lens plane and $\mathbf{\eta}$ is the position vector of the source in the source plane.
In this notation, time delay functional reads:
\begin{equation}
t_d(\mathbf{x},\mathbf{y})=\frac{D_sR^2_E(1+z_l)}{D_lD_{ls}}\left[\frac{1}{2}|\mathbf{x}-\mathbf{y}|^2-\psi(\mathbf{x})+\phi_m(\mathbf{y})\right],
\end{equation}
where $\psi(\mathbf{x})$ is dimensionless deflection potential and $\phi_m(\mathbf{y})$ is chosen so that the minimum arrival time is zero.
Since there is a lot of compact structure in galaxies (stars, black holes, compact dark matter clumps), we consider the lens  described by a point mass, with the  lensing potential $\psi(\mathbf{x})=\ln{x}$ and  $\phi_m(\mathbf{y}) = 0.5\;(x_m - y)^2 - \ln{x_m}$ with $ x_m = 0.5 \;(y + \sqrt{y^2 + 4})$.

For universality, we define the dimensionless parameter
$w=8\pi M_{Lz}f$ which serves as a comparison between the barrier and the wavelength. At this point we would like to be as general as possible, hence we introduce the redshifted mass $M_{Lz}=M(1+z_l)$. Of course, for nearby sources one should neglect cosmological effects (setting $z_l = z_s = 0$).

For the case in the limit of $w >> 1 $, Eq.\ref{amplitude} simplifies to:
\begin{equation}
|F(f)|=\sqrt{|\mu_+|+|\mu_-|+2|\mu_+\mu_-|^{1/2}\sin{(2\pi f\Delta t_d)}},\label{geo}
\end{equation}
where $|\mu_+|$ and $|\mu_-|$ are the magnifications of the brighter and fainter image in the geometrical optics limit.
$\Delta t_d$ is the time delays between multi-signals. For the point mass lens,
\begin{equation}
\Delta t_d=4M_{Lz}\left[\frac{y\sqrt{y^2+4}}{2}+ln\left(\frac{\sqrt{y^2+4}+y}{\sqrt{y^2+4}-y}\right)\right]\sim8M_{Lz}y.
\end{equation}
The last term in Eq.\ref{geo} is the interference between images.
For the light sources whose multiple images are independent or the transient GW sources whose lasting times are smaller than $\Delta t_d$, this term vanishes.
We will see two separate signals amplified by $\mu_{\pm}$ arriving at different times.
For the continuous GWs considered in this work, the term gives the interference modulation.

In the earlier works \citep{Jung2017,Christian2018,Takahashi2003} mentioned above, the authors studied distortions of the waveform of a chirp signal whose frequency increases with time, assuming fixed $y$ since the chirping time scale is very small.
On the contrary, we consider the sources of continuous GW whose frequencies can be assumed constant for a long time.

At last, it is worth stressing that, while the fringes of light are determined by $|F|^2$, i.e. the intensity,
in the context of GWs the amplitude $|F|$ is relevant.

\section{Peculiar motions and detectability of the fringes}
Fig.\ref{point} shows the amplification factor as a function of the source position $y$ expressed as a fraction of the Einstein radius of the lens for $w=1, 3, 9, 300$ in the case of point mass, respectively.
One can see the diffraction and interference fringes and develop the idea of their detectability.

\begin{figure}
 \includegraphics[width=8cm,angle=0]{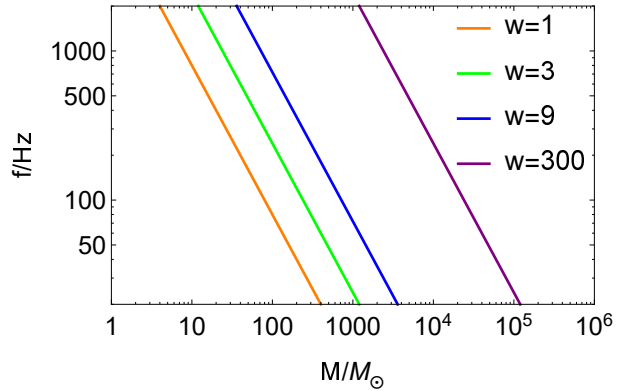}
  \caption{Relation between the GW frequency and the lens mass for different values of $w$.
  }\label{mf}
\end{figure}

First of all $w$ should not be too small, since the amplitude of diffraction fringes would not be detectable in that case. We propose the $w > 1$ criterion as reasonable. Recalling  the definition of $w$ one can see that for the frequencies of interest e.g. of order of, $f=1\;kHz$ masses of the lenses satisfying $w > (1, 3, 9, 300) $ criterion would be
$M/M_{\odot} > (8.1, 24.3, 72.9, 2430)$.
Fig.\ref{mf} shows the dependence between $w$, the lens mass and the GW frequency. Then, one can see that the amplitude pattern of fringes is damped. However, they are noticeable beyond the Einstein ring radius when $y<3$. This is an interesting new feature of GW diffraction effects, different from the standard strong lensing/microlensing case.

At last, in order to see a complete diffraction or interference fringe, the monitoring time should be sufficiently large $\Delta t>t_f$, where $t_f$ is the fringe
time scale, such that the peculiar motions would result in relative spatial change (see more details later). Appropriate timescale for transit of fringes is $t_f = t_E /w$, where $t_E$ is the Einstein radius crossing time.
This was noticed in \citep{Naderi2018} who discussed detectability of primordial black holes by detecting the diffraction patterns in lensed quasars. In a similar way, \citep{Mehrabi2013} studied the wave optics features of gravitational microlensing by a binary lens composed of a planet and a parent star. These papers discussed lensing of light, while \citep{Ruffa1999}  formulated an idea of the Milky Way's own supermassive black hole amplifying the continuous wave signal from an extragalactic rotating neutron star and the resulting diffraction pattern.

We summarize the criteria of detectability of the fringes for a point mass lens:
\begin{itemize}
\item  $w>1$ to have enough amplification variation;
\item  $y<3$ to detect fringes before they are damped;
\item  $\Delta t>t_f$ to see a fringe pattern.
\end{itemize}

\begin{figure*}
 \centering
 \includegraphics[width=5cm,angle=0]{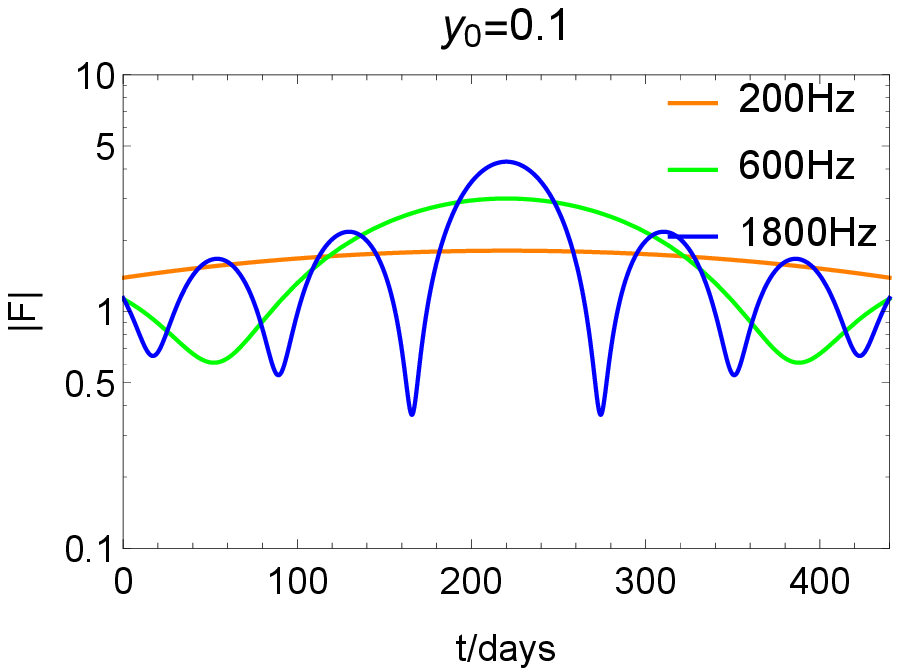}
 \includegraphics[width=5cm,angle=0]{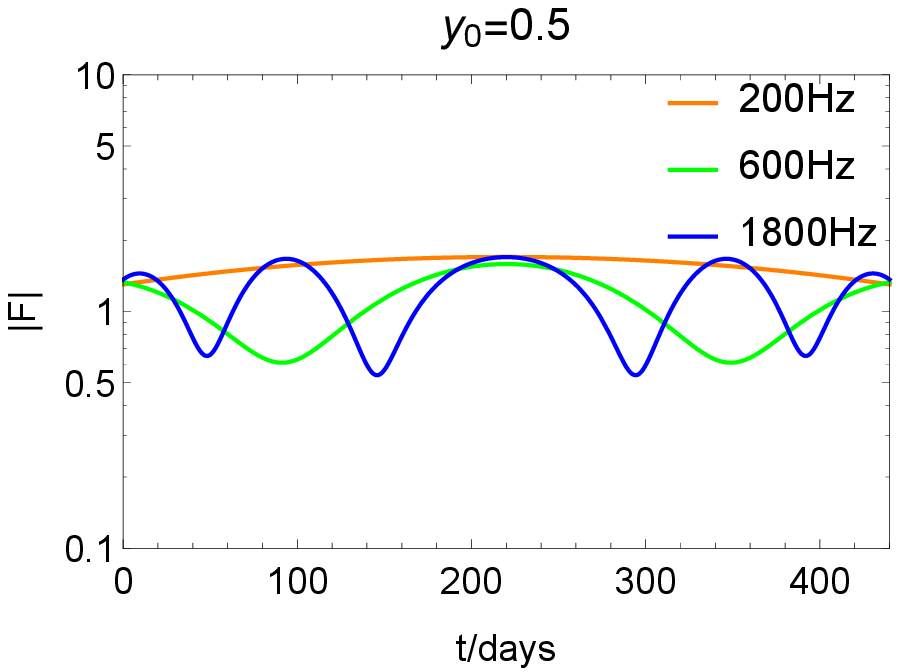}
 \includegraphics[width=5cm,angle=0]{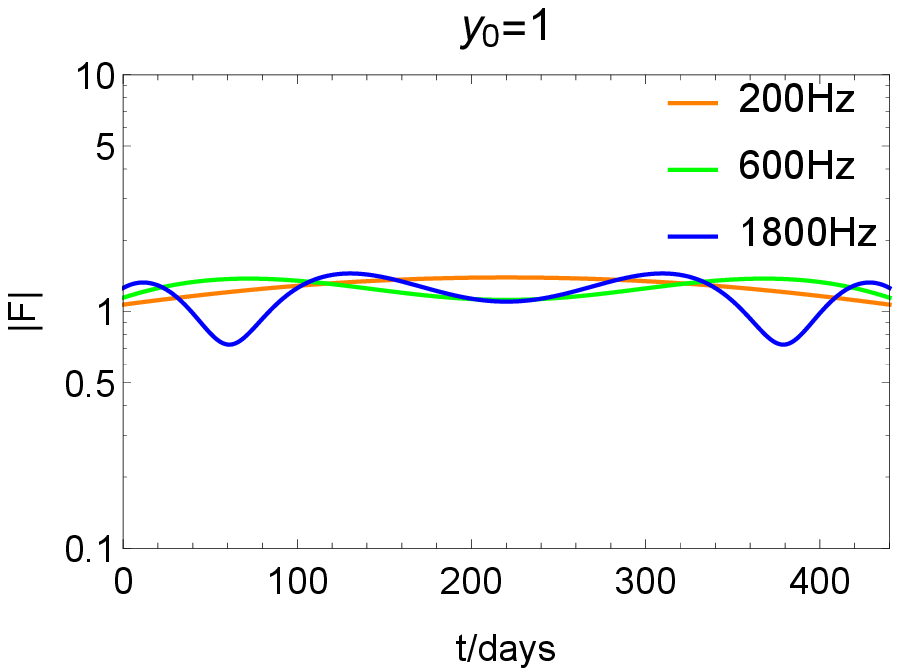}
  \caption{ Amplitude variation of a monochromatic GW in time interval  corresponding to two Einstein crossing times for a $ 40\;M_{\odot}$ lens. Closest approach time corresponds to $t_E$, diffraction patterns are shown for three frequencies of the GW
$f=200Hz,600Hz,1800Hz$. Panels from left to right correspond to three values of the source position at the closest encounter $y_0=0.1,0.5,1$, respectively.}\label{star}
\end{figure*}

\begin{figure*}
 \centering
 \includegraphics[width=5cm,angle=0]{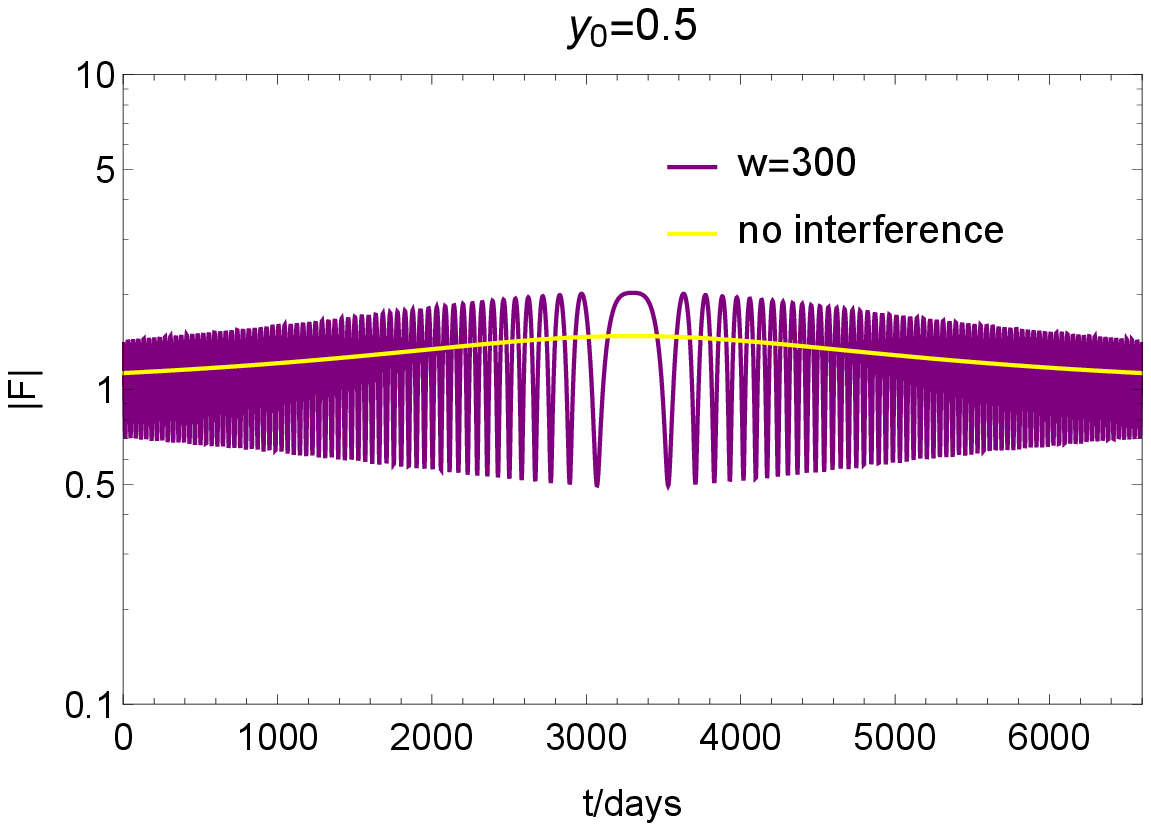}
 \includegraphics[width=5cm,angle=0]{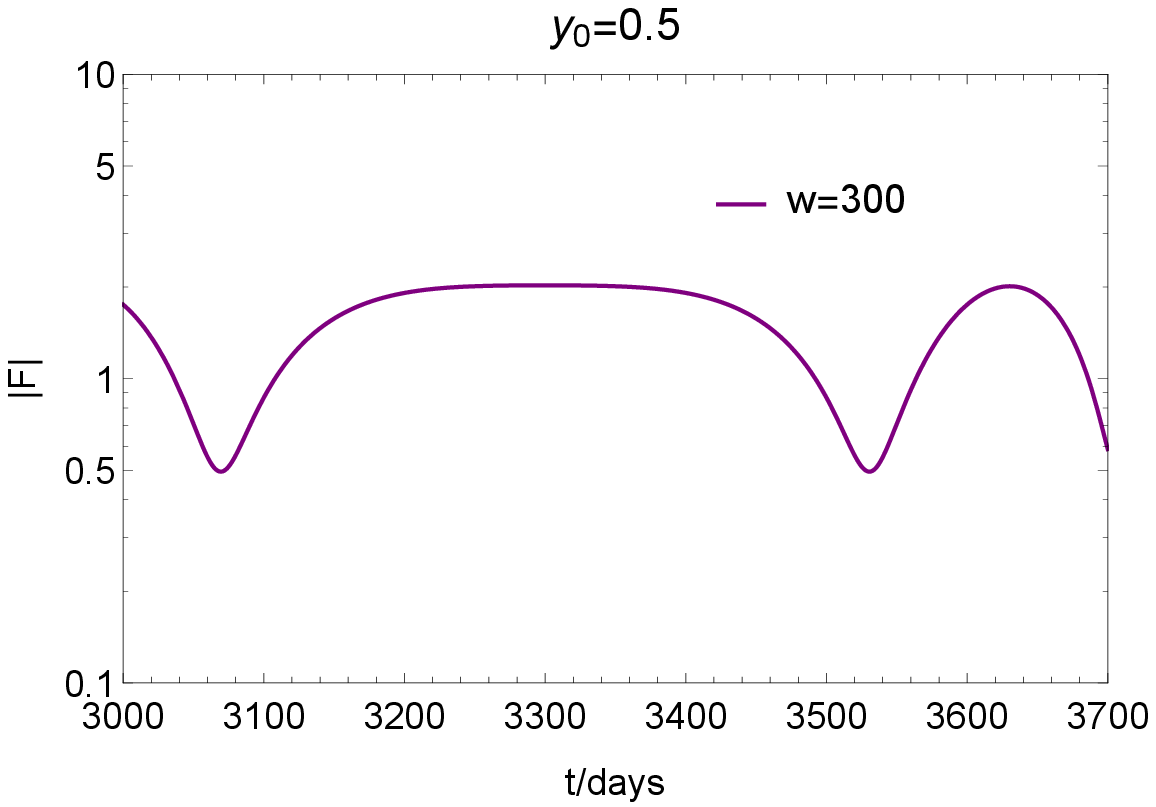}
 \includegraphics[width=5cm,angle=0]{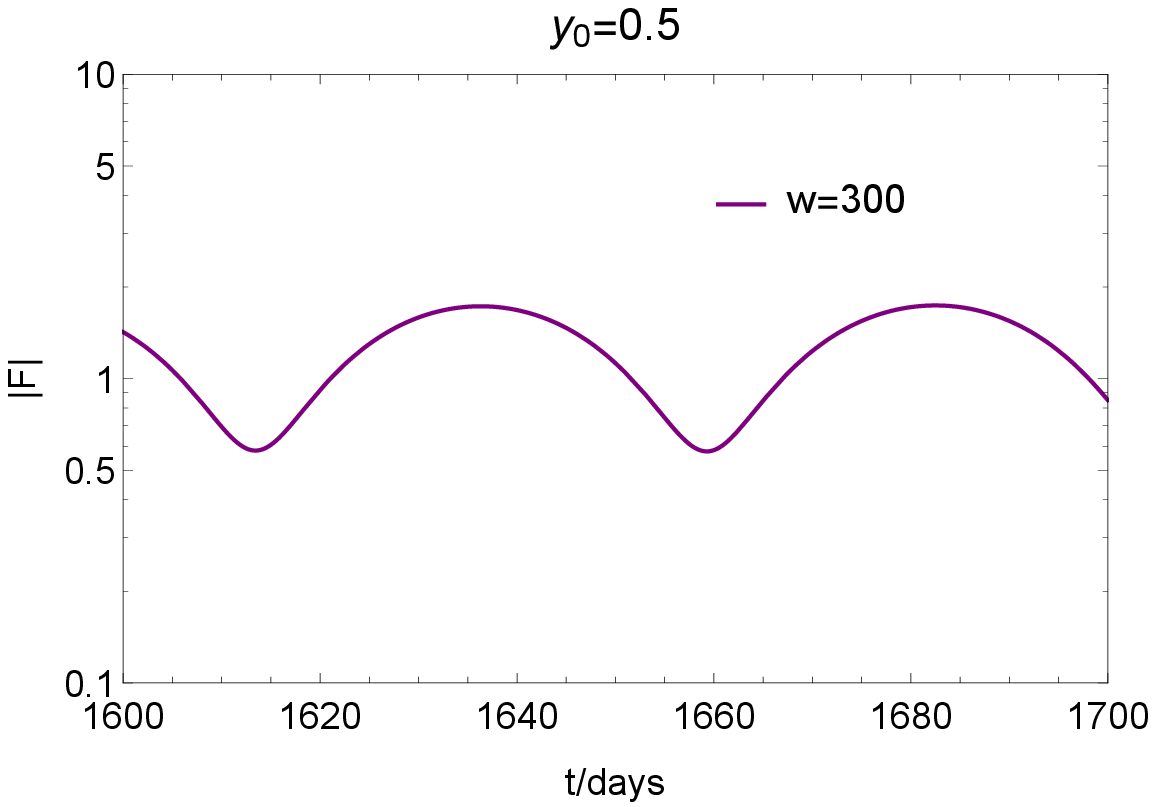}
  \caption{Amplitude variation of a monochromatic GW in time interval  corresponding to two Einstein crossing times for a $ 10^4\;M_{\odot}$ lens in the geometric optics limit $w=300$. Source position at the closest encounter is $y_0 = 0.5$. For comparison, the yellow line shows the corresponding amplification of light from unresolved lensed images.
Middle and right panels enlarge the intervals of $700$ days and $100$ days
around the closest encounter and away from it, respectively.
  }\label{DM}
\end{figure*}

Of course, with GW signals one is not able to observe the fringe patterns on the screen. However, peculiar motions (with respect e.g to the CMB frame) of the components of lensing system: the observer $\mathbf{v}_{obs}$, the lens $\mathbf{v}_{l}$ and the source $\mathbf{v}_{s}$ can be combined to the effective velocity of the source with respect to the lens \citep{Kayser}:
\begin{equation}
\mathbf{v}_{eff} = \mathbf{v}_{s} - \frac{1+z_s}{1+z_l} \frac{D_s}{D_l}\mathbf{v}_{l} + \frac{1+z_s}{1+z_l} \frac{D_{ls}}{D_l} \mathbf{v}_{obs}.
\end{equation}
Using this effective relative velocity in the system, $t_f=\frac{R_E}{v_{eff}w}$ provides the time scale of observability of diffraction fringes.
As the source position changes with respect to the lens according to:
$y(t) = \sqrt{y_0^2 + (\frac{t - t_0}{t_E})^2} $, where $t_0 $ is the time of the closest encounter occurring at the source position $y_0$,
the amplitude of GW from a periodic source gets modulated.
This creates opportunity to see the diffraction fringes with a favorable configuration of the source and the lens.

Central to our idea is that the source of GW is continuous, with the frequency not significantly changing during the observation time.
For LIGO, and the third generation of ground based detectors the source of periodic gravitational waves is expected to be isolated neutron stars emitting at
frequencies $f\sim 10 - 10^3\; Hz $. In the case of galactic sources and lenses, distances can be regarded as Euclidean distances and the Einstein time reads:

\begin{equation}
\begin{split}
t_E \approx 34.7 \;day \; \sqrt{4 \frac{D_l}{D_s} \left( 1- \frac{D_l}{D_s} \right) } \left( \frac{D_s}{8 \; kpc} \right)^{1/2}\\
 \times\left(\frac{M}{M_{\odot}} \right)^{1/2} \left(\frac{v_{eff}}{200\;km/s}\right)^{-1},
\end{split}
\end{equation}
which would be 220 days for a $ 40\;M_{\odot}$ lens.
One can see from Fig.\ref{mf} that in such case for a source emitting GWs with frequency $f = 600\;Hz$, $w$ parameter will be $w=3$ and $t_f=73.3$ days.
This is advantageous because this kind of amplitude modulation can be distinguished from the modulation caused by Earth rotation or orbital motion.
In Fig.\ref{star} we show the pattern of amplitude modulation of sources emitting at frequencies  $200\;Hz$, $600\;Hz$ and $1800\;Hz$ lensed by a $ 40\;M_{\odot}$ point mass. This corresponds to $w$ parameter: $w=1,3,9$ respectively.
In the case of much more massive lens like a compact dark matter clump or a black hole, one could have $w \gg 1$ corresponding to the geometric optics limit. Yet, the amplitude of a monochromatic GW can be modulated due to interference between two image signals (see Eq.\ref{geo}). For example, if the lens has mass $M=10^4\;M_{\odot}$, Einstein time would be $t_E=3300$ days and for a $f=240\;Hz$ source $w$ parameter would be $w=300$, $t_f=11$ days. We display the results in Fig.\ref{DM}.

\section{Observing feasiblity}

A question arises, how probable is to observe microlensed monochromatic GWs probing the fringes as discussed above.
In the context of ground based detectors, the most realistic source of continuous GW radiation are spinning NSs.
Extrapolating the NS birthrate \citep{Narayan1990} and
from the number of supernovae required to account for
the heavy element abundances in the Milky Way \citep{Arnett1989}
one may expect that the galactic bulge contains about $10^9$ NSs.
Globular clusters are expected to contain $10^3$ NSs \citep{Grindlay}, which could be a target population for GW detectors.
Putting aside the problem how many of them would be accessible to the third generation of detectors, as a first order estimate one may use the classical approach to the microlensing probability used in optical studies. In this section we reintroduce the dimensioned fundamental constants $G$ and $c$ in the formulae.

The optical depth is the probability that a given source falls into the Einstein radius of any lensing star along the line of sight:
\begin{equation}
\tau = \int_0^{D_s} n(D_l)\pi R_E^2 dD_l.
\end{equation}
Let us recall, at this point, that unlike in the optical microlensing,
wave effects discussed by us can be observed up to the source position $y_{max}=3$ Einstein radii. Therefore optical depth can be boosted by a factor
of $y_{max}^2$. Of course the optical depth for lensing is scenario dependent. As a first scenario, let us consider sources located in the bulge, lensed by stellar mass lenses located in the Galactic disk.
An order of magnitude estimate is given in the original papers  \citep{Paczynski1986, Paczynski1991}. Under assumption of constant density of lenses along the line of sight, one has:
\begin{equation}
\tau = \frac{1}{2 c^2} \frac{G M(<D_s)}{D_s} = \frac{v_{rot}^2}{2 c^2}.
\end{equation}
With the Milky Way's bulge as a target, $v_{rot} \approx 220 \;km/s$ and
$\tau \approx 2.7\times 10^{-7}$.
In our case, this should be modified to
\begin{equation}
\tau_{GW} \sim  f_l y_{max}^2 \times 10^{-7} \sim f_l \times 10^{-6},
\end{equation}
where $f_l$ is the fraction of lens mass larger than solar mass since smaller mass will enter the wave optics limit. Of course, more realistic estimates have to consider proper mass distribution of potential lenses (e.g. exponential) and proper models of stellar systems considered (e.g. double exponetntial disk, spheroidal bulge, Plummer sphere model of globular clusters, etc.).
Probability of GW microlensing events expected in a time interval $\Delta t$ can be assessed as:
\begin{equation}
P \sim \tau_{GW} \frac{\Delta t}{t_f}
\end{equation}
where the time scale of GW wave effects is $t_f = t_E/w$.
Considering the ET operation time as $\Delta t=10\; yr$, one has:
$P \sim f_l (t_f / 1\;month)^{-1} \times 10^{-4}$. This means that if only a fraction $f_s = 10^{-5}$ of the source population of $n = 10^9$ bulge NSs would be accessible to the ET, such GW microlensing events are expected to occur.
Let us emphasize that unlike in the optical, one does not face the problem of source resolution in the crowded field or light blending.

Another, even more promising scenario is associated with lenses located in rich globular clusters like M22 seen against the Galactic bulge. A typical velocity dispersion in a globular cluster is $\sigma \sim 10\; km/s$,
so the typical transverse velocity of the lens is well approximated by that of a cluster as a whole $\sim 200 \; km/s$. As shown by \citep{Paczynski1994}
the optical depth for microlensing in such case is given by:
\begin{equation}
\tau = \frac{\sigma^2}{c^2} \frac{2 \pi}{\phi} \left( 1 - \frac{D_l}{D_s} \right) \approx 2.4 \;10^{-5} \left( \frac{\sigma}{10\; km/s} \right)^2 \left(\frac{1'}{\phi} \right)
\end{equation}
where the last equality holds when $D_l / D_s << 1$, and $\phi$ denotes the angular distance from the cluster center. In the optical this implies that
looking at the target field in the bulge
within a few arcminutes of the cluster center the lensing is dominated by
objects associated with the cluster itself. This observation triggered much interest since knowledge of the distance and transversal velocity of lenses belonging to the cluster may result with accurate estimates of their masses.
First confirmed microlensing of the bulge star at $D_s = 8.2\;kpc$ by a low mass object in the globular cluster M22 located at $D_l = 2.6\;kpc$ has been reported in \citep{Sahu, Pietrukowicz2012}.
In the context of our considerations, this scenario predicts the optical depth
\begin{equation}
\tau_{GW} \sim  f_l \; y_{max}^2\; \left(1/\phi \right) \times 10^{-5} \sim f_l \; (1/\phi) \times 10^{-4}
\end{equation}
where, $\phi$ is expressed in arcminutes. Probability of GW microlensing events in such scenario is: $P \sim f_l \; (1/\phi) (\Delta t / 10\;yr ) (t_{GW} / 1\;month)^{-1} \times 10^{-2}$. This is particularly promising scenario since the GW signal from the bulge NS can pass through the very center of the globular cluster with very high probability of being lensed. Moreover, it can be lensed by massive objects expected to reside in the center of the cluster, including the intermediate mass black holes ($M \sim 10^2 - 10^4 \;M_{\odot}$).

At last, in \citep{Safonova} the authors considered microlensing scenario where both the source and the lens belong to the globular cluster. They estimated the optical depth for such events as: $\tau \approx 10^{-3}$. Considering that globular clusters anchor about $n=10^3$ NSs and are much closer than the Galactic bulge this is also a promising scenario from our perspective.

\section{Conclusion and discussions}
We proposed an observational strategy to test the wave nature of GWs by monitoring the amplitude modulation of lensed sources of continuous gravitational signals. One of the candidates of continuous GW signal are spinning neutron stars slightly deformed from perfect spherical symmetry, which are targets of ground based detectors. Their detection is fairly difficult, since the signal is
strongly modulated by the Earth's rotation and orbital motion. Moreover this modulation is different for every sky position. Diffraction and interference fringes caused by intervening mass acting as a lens and moving with respect to the source also produce modulation. Luckily the time scales of these effects are significantly different from time scales involved with the motion of the detector.
In the case of monochromatic GWs search, strategies are different from chirping signals of coalescing binaries. Depending on what is a priori known about sources of GWs, searches can be targeted (known position and GW frequency, e.g. Crab pulsar), directed (only the sky position known), and all-sky (or blind) searches. In the setting discussed by us only blind searches or directed searches (incase of globular cluster lensing) can be applied.
The GW signal coming from a rotating NS is so weak that to detect it in the detector's noise one has to analyze months-long segments of data. Fully coherent analysis of such amount of data is computationally
prohibitive in the case of all-sky searches \citep{Brady1998, Brady2000}.  Therefore different hierarchical two-stage schemes were developed, where in the first stage shorter segments of data are analyzed coherently and then in the second stage the results are combined in an incoherent way.
Implementation of the {\cal {F}}-statistic \citep{Jaranowski1998} consists in
coherent search over two-day periods, followed by a search for coincidences among the candidates from the two-day segments. The time scales for fringe modulation discussed here, are larger and one can expect that respective amplitude modulation could be detectable with current techniques. 

In this work, we suggested plausible configurations comprising bulge NSs lensed by massive objects in the disk, or lensed by massive objects in the globular clusters seen against the bulge or globular cluster NSs lensed by massive stars in the same cluster. More detailed elaboration of these scenarios for the further study.
This is worth pursuing since the verification of spatial diffraction fringes of lensed continuous gravitational sources would strengthen our understanding of GWs and contribute to setting up the quantum gravity and shed light on the nature of the
graviton. We are looking forward to seeing this phenomenon observed by second-generation detectors like advanced LIGO and Virgo, by the third-generation detectors like the Einstein Telescope or space-born detectors like LISA.

\section*{Acknowledgments}
KL was supported by the National Natural Science Foundation of China (NSFC) No. 11603015
and the Fundamental Research Funds for the Central Universities (WUT:2018IB012).
MB was supported by the Key Foreign Expert Program for the Central Universities No. X2018002.
XF was supported by the NSFC 11673008 and Newton
International Fellowship Alumni Follow on Funding.

\clearpage

\end{document}